\documentclass[twocolumn]{aastex62}

\graphicspath{{./}{figures/}}
\usepackage{bm}
\usepackage{amsmath}

\newcommand{\dd}{\mathrm{d}}
\begin{document}

\title{Self-Regulated Cosmic-Ray Electron Acceleration in High-Mach-Number MHD-PIC Shocks}
\correspondingauthor{Alisa Galishnikova}
\email{astro.alisag@gmail.com}

\author[0000-0002-6300-3191]{Alisa Galishnikova}
\affil{Center for Computational Astrophysics, Flatiron Institute, 162 Fifth Avenue, New York, NY 10010, USA}
\affil{Department of Astrophysical Sciences, Princeton University, 4 Ivy Lane, Princeton, NJ 08544, USA}

\author[0000-0002-1030-8012]{Siddhartha Gupta}
\affil{Department of Astrophysical Sciences, Princeton University, 4 Ivy Lane, Princeton, NJ 08544, USA}
\affil{Saha Institute of Nuclear Physics, 1/AF, Bidhannagar (Salt Lake City), Kolkata 700064, India}

\author[0000-0001-9179-9054]{Anatoly Spitkovsky}
\affil{Department of Astrophysical Sciences, Princeton University, 4 Ivy Lane, Princeton, NJ 08544, USA}

\author[0009-0009-7676-6188]{Xiaochen Sun}
\affil{Department of Astrophysical Sciences, Princeton University, 4 Ivy Lane, Princeton, NJ 08544, USA}
\affil{Department of Astronomy and Columbia Astrophysics Laboratory, Columbia University, New York, NY 10027, USA}

\begin{abstract}
Shock waves in supernova remnants are thought to accelerate cosmic rays (CRs) via diffusive shock acceleration (DSA). If acceleration is efficient, CR-modified shock models and nonlinear DSA predict smoothly decelerated precursors and concave spectra that harden with energy. These predictions are in tension with observations, which favor spectra closer to the test-particle form, $4 \pi f(p) p^{4} = {\rm const}$ or steeper. We use one- and two-dimensional magnetohydrodynamic-particle-in-cell simulations to follow the long-term nonlinear coupling of the thermal plasma and CR electrons in non-relativistic quasi-parallel, high-Mach-number shocks. We choose electrons over ions for their larger charge per unit mass, probing the limits of current-driven feedback. We vary the supplied CR-electron mass fraction to measure the shock's response. Our injection model determines the effective injection and acceleration from particle trajectories through the evolving upstream turbulence, rather than a prescribed shock injection fraction. 
The reflected CR-electron current excites Bell-type non-resonant turbulence. The resulting magnetic structures scatter and isotropize the current-carrying particles, limiting the coherent upstream current and suppressing the formation of a spatially extended CR-pressure precursor. Across the CR-electron mass supply explored, the saturated upstream magnetic energy and far-upstream CR current converge to similar levels. In 1D, the downstream high-energy spectrum approaches the test-particle DSA shape across the resolved momentum range, with nonthermal pressure reaching a few tens of percent of the upstream ram pressure. 
These results suggest that, once injection is sufficient to sustain strong current-driven turbulence, supernova-remnant shocks can approach a common regulated state largely insensitive to the details of CR injection.
\end{abstract}


\section{Introduction} \label{sec:intro}
Shock waves in supernova remnants (SNRs) are thought to produce non-thermal charged particles, cosmic rays (CRs), through first-order Fermi acceleration, or diffusive shock acceleration \cite[DSA; e.g.,][]{Krymskii1977DoSSR.234.1306K,Axford1977ICRC...11..132A,Bell1978MNRAS.182..147B,BlandfordOstriker1978ApJ...221L..29B}. In the test-particle limit of DSA, the isotropic momentum distribution of CRs, $f(p)$, in a strong (highly-supersonic) shock is expected to follow  $f(p)\propto p^{-4}$, where $p$ is the magnitude of the particle momentum. This prediction is widely used to interpret nonthermal emission from SNRs. However, if SNR shocks convert a substantial fraction of upstream kinetic energy into the nonthermal particles, particles cannot remain passive -- they must eventually feed back on the shock structure. 

Classical CR-modified shock models and semi-analytic nonlinear DSA (NLDSA) describe this feedback through the CR pressure gradient that exerts a force on the inflowing plasma. This forms a smooth precursor in which the flow decelerates before reaching the gas subshock, thereby reducing the effective density compression experienced by low energy particles near the shock \citep[e.g.,][]{Blandford1980ApJ...238..410B, Heavens1983MNRAS.204..699H, BlandfordEichler1987PhR...154....1B, Drury1981ApJ...248..344D, MalkovDrury2001RPPh...64..429M, Blasi2004APh....21...45B, Amato2006MNRAS.371.1251A}. Since higher-energy particles have larger diffusion lengths, they sample a larger effective compression ratio than lower-energy particles. This produces a concave spectrum that hardens with energy. Such predictions are in tension with observations, which often favor steeper spectra, closer to or even steeper than the test-particle DSA expectation \citep[e.g.,][]{Fermi2016ApJS..224....8A}.
In hybrid simulations (where protons are treated as particles and electrons as a fluid), efficient acceleration produces spectra close to the test-particle limit, despite substantial magnetic-field amplification. This behavior has been interpreted in terms of a ``postcursor,'' in which downstream magnetic structures drift relative to the thermal plasma and increase the effective downstream advection of CRs \citep{HaggertyCaprioli2020ApJ...905....1H,DiesingCaprioli2021ApJ...922....1D}. In that picture, the spectral modification is controlled primarily by the scattering-center velocity downstream of the shock. 

In high-Mach-number quasi-parallel shocks, the shock-reflected CR population carries an upstream current, compensated by a return current in the background plasma, which drives the non-resonant Bell instability \cite[]{Bell2004MNRAS}. The streaming CRs, therefore, scatter off turbulence they generate themselves. In CR-modified shock treatments, particle transport is usually described by an effective diffusion coefficient, $\kappa(p,x)$, acting on a close-to-isotropic CR distribution \citep[e.g.,][]{Blasi2004APh....21...45B}. However, the upstream CR flux and the scattering that controls it are coupled: streaming particles drive magnetic turbulence, which in turn alters their transport and the current driving further growth. 
Models incorporating self-generated turbulence describe this feedback through a closure relating magnetic fluctuations to particle diffusion \citep{Amato2006MNRAS.371.1251A}. In the non-linear Bell turbulence this feedback may provide a self-regulation mechanism: stronger currents drive stronger turbulence, but stronger turbulence removes the anisotropy that drives it. {This raises the possibility that, once the particle supply is sufficient to sustain strong current-driven turbulence, shocks with different CR injection conditions may approach a common regulated state. Such convergence would reduce the sensitivity of magnetic amplification and the CR population to the details of injection.}

Fully kinetic particle-in-cell (PIC) simulations provide a self-consistent way to study these processes, but they are computationally expensive for following shock waves over macroscopic length scales and long dynamical times while simultaneously resolving both electron and ion kinetic scales. In this work, we use the MHD-PIC method \citep{Bai_2015}, in which the thermal electron-ion plasma is evolved as a magnetohydrodynamic (MHD) fluid, while cosmic ray electrons (CRe) are advanced kinetically and exchange momentum and energy with the fluid. This approach does not resolve the microphysics of thermal injection, but it captures the long-term coupling between CR currents, self-generated turbulence, and the MHD shock structure. 

{We use this method to study how high-Mach-number quasi-parallel shocks respond to changes in the supplied CRe population, and whether CR feedback produces an extended smooth precursor.} We focus on CR electrons because their larger charge per unit rest mass allows a stronger driving current than ions at the same CR mass loading and streaming velocity. This provides an extreme case for testing how strongly current-driven electromagnetic feedback can modify the shock. We find that the coupled evolution of the CR current and Bell-type turbulence regulates the shock structure. Future work will test whether the same regulation operates for CR ions.

We initialize a prescribed population of preheated electrons upstream as test particles, which move in the MHD fields without exerting feedback on the fluid. Their momentum and energy feedback is activated according to a positional criterion near the shock, described in \S2.2. The fraction of this seed population that reflects upstream and enters the acceleration cycle is determined by integrating particle trajectories in the evolving CR-driven upstream turbulence. Thus, although the upstream particle supply is prescribed, the effective injection into acceleration responds to the evolving shock and turbulence. The coupled evolution of particles and fields allows us to follow the long-term feedback between particle acceleration, upstream streaming, and self-generated turbulence.


This paper is organized as follows. In \S2 we describe the MHD-PIC method, the numerical setup, and particle initialization and the criterion for activating particle feedback. In \S3 we present the 1D simulations, showing Bell-driven magnetic amplification, current saturation, weak hydrodynamic modification, and the resulting particle spectra. We then extend the analysis to 2D and discuss convergence tests. In \S4 we interpret the results in terms of Bell-regulated current saturation and compare them with classical CR-modified shock and NLDSA expectations. We summarize our conclusions in \S5.

\section{Method} \label{sec:method}
\subsection{MHD-PIC method}
We use the MHD-PIC framework in Athena to evolve the coupled dynamics of the thermal plasma and CR particles. We briefly summarize the method here (see \citealt{Bai_2015} for details). The thermal plasma is evolved as an ideal MHD fluid, while CRs are represented by kinetic macroparticles that exchange momentum and energy with the MHD fields. Each CR macroparticle of species $s$ obeys the Lorentz-force equation,
\begin{equation}
    \frac{\dd \gamma_s {\bm v}_s}{\dd t} = \left( \frac{q}{m c}\right)_s \left( c\bm{\mathcal E} + {\bm v}_s \times { \bm B}\right)
\end{equation}
where ${\bm v}_s$ is the particle 3-velocity, $q_s$ and $m_s$ are its charge and mass, and $\gamma_s = \left(1-\frac{v_s^2}{c_L^2}\right)^{-1/2}$ is the Lorentz factor. Here, $c_L$ is an (artificial) speed of light used in the particle pusher; $c_L$ is chosen to be much larger than the characteristic fluid and MHD wave speeds.
The electric field $\bm{\mathcal E}$ and magnetic field ${\bm B}$ are taken from the MHD fluid. For ideal MHD,
\begin{equation}
    \bm{\mathcal E} = -\frac{{\bm v}_g \times {\bm B}}{c}.
\end{equation}
Extensions can include CR-Hall or resistive terms; see \citealt{Bai_2015} for implementation details.

An ensemble of CRs with bulk velocity ${\bm v}_{\rm CR}$ and number density $n_{\rm CR}$ carries charge density $\rho_{\rm CR}\equiv q\,n_{\rm CR}$ and current density ${\bm J}_{\rm CR}=\rho_{\rm CR}{\bm v}_{\rm CR}$. The corresponding Lorentz force density on the CRs is
\begin{equation}
	{\bm F}_{\rm CR} = \rho_{\rm CR}\bm{\mathcal E} + \frac{1}{c}\,{\bm J}_{\rm CR}\times{\bm B}.
\end{equation}
The total current is the sum of gas and CR contributions,
\begin{equation}
	{\bm J}_{\rm tot} = {\bm J}_{\rm gas}+{\bm J}_{\rm CR}
	= q_i n_i {\bm v}_i + q_e n_e {\bm v}_e + \rho_{\rm CR}{\bm v}_{\rm CR},
\end{equation}
where $q_i=+e$ and $q_e=-e$, and ${\bm v}_i$ and ${\bm v}_e$ are the mean ion and electron velocities. We assume a quasi-neutral plasma,
\begin{equation}
    \rho_{\rm tot} = q_i n_i + q_e n_e + \rho_{\rm CR} \approx 0,
    \ |\rho_{\rm CR}|\ll |q_i n_i|\sim |q_e n_e|.
\end{equation}

CR backreaction enters the MHD equations as source terms in momentum and energy: $-{\bm F}_{\rm CR}$ and $-{\bm J}_{\rm CR}\cdot\bm{\mathcal E}$. Using ${\bm J}_{\rm CR}=\rho_{\rm CR}{\bm v}_{\rm CR}$ and ${\bm v}_{\rm CR}\cdot({\bm v}_{\rm CR}\times{\bm B})=0$, one may write
\begin{align}
    &\frac{\partial \rho}{\partial t} + \nabla \cdot (\rho {\bm v}_g) = 0,\\
	&\frac{\partial (\rho {\bm v}_g)}{\partial t} + \nabla \cdot \left[\rho {\bm v}_g {\bm v}_g + \left(P_g + \frac{B^2}{8\pi}\right){\bm I} - \frac{{\bm B}{\bm B}}{4\pi}\right] \notag \\
    & \qquad = -{\bm F}_{\rm CR},\\
    &\frac{\partial E}{\partial t} + \nabla \cdot \left[\left(E+P_g+\frac{B^2}{8\pi}\right){\bm v}_g - \frac{({\bm B}\cdot{\bm v}_g){\bm B}}{4\pi}\right] \notag \\
    & \qquad = -{\bm v}_{\rm CR}\cdot{\bm F}_{\rm CR},
\end{align}
where $\rho$, $P_g$, and ${\bm v}_g$ are the gas density, thermal pressure, and velocity, and ${\bm I}$ is the identity tensor. The total energy density (gas plus magnetic) is
\begin{equation}
    E \equiv \frac{P_g}{\gamma-1} + \frac{1}{2}\rho v_g^2 + \frac{B^2}{8\pi}.
\end{equation}
The induction equation is
\begin{equation}
	\frac{\partial {\bm B}}{\partial t} = -c\,\nabla\times\bm{\mathcal E}.
\end{equation}
In {\tt Athena} \citep{Stone2008ApJS..178..137S}, the particle push and MHD update are coupled using a second-order predictor-corrector scheme.

\subsection{Setup}\label{sec:setup}
We initialize a magnetized, quasi-parallel shock by driving a super-Alfv\'enic and super-sonic flow into a perfectly conducting and reflecting wall at the left boundary ($x\equiv 0$). All simulations are performed in the downstream rest frame, so the upstream plasma propagates leftward with velocity $-v_0\,\hat{\bm x}$.

We adopt dimensionless units with the factor of $4\pi$ absorbed into $B^2$. Lengths are normalized to the ion skin depth, $d_i\equiv c/\omega_{pi}$, where $c$ is the speed of light and $\omega_{pi}$ is the ion plasma frequency. Times and velocities are normalized to $\Omega_{ci}^{-1}$ and $v_{\rm A,0}$, respectively, where $\Omega_{ci}\equiv(q/mc)_i B_0$ is the ion cyclotron frequency and $v_{\rm A,0}\equiv B_0/\sqrt{\rho_0}$ is the upstream Alfv\'en speed. The species parameter $(q/mc)_s$ is normalized to its background-ion value. In these code units, $\rho_0=1$, $B_0=1$, and $(q/mc)_i=1$.

We study a supersonic and super-Alfv\'enic flow with
\begin{equation}
    M_{\rm A}\equiv \frac{v_0}{v_{\rm A,0}}=50, \qquad
    M_s \equiv \frac{v_0}{v_{{\rm th},i,0}}=50 .
\end{equation}
For the particle pusher we use an artificial speed of light $c_L=10^4\,v_{\rm A,0}$, corresponding to $c_L=200\,v_0$ for the fiducial runs, which is much larger than all MHD speeds in our main runs and keeps the bulk of the CR particles non-relativistic for most of the duration of the runs.
The initial magnetic field is uniform and inclined by $\theta=30^\circ$ relative to $\hat{\bf x}$, so that $B_{x,0}=B_0\cos\theta$ and $B_{y,0}=B_0\sin\theta$ (we have also performed simulations with an inclination of $20^\circ$ and found that all results of the paper still hold).

For a strong shock with $\gamma=5/3$, the Rankine-Hugoniot jump conditions imply a compression ratio $\simeq 4$, so in the downstream frame the ideal shock front propagates at
\begin{equation}
    {\bm v}_{\rm sh}=\frac{v_0}{3}\,\hat{\bm x}.
\end{equation}
Throughout the paper, we normalize energies to $E_{\rm sh}\equiv \frac{1}{2} m v_{0}^2$ and momenta to $p_0\equiv m_e v_0$.

We adopt a reduced ion-to-electron mass ratio $m_i/m_e=100$, corresponding to $(q/mc)_e=-100$ for CR electrons (CRe) in our code units. 
Particles are introduced upstream as test particles, i.e., they are advanced in the MHD fields but do not contribute to the CR feedback terms in the MHD update. At each time step, test particles are injected in the far upstream in a slab of thickness $\simeq 100 d_i$ centered $1000d_i$ ahead of the anticipated shock position $x_{\rm sh,exp}(t)=v_{\rm sh, 0}t$, where $v_{\rm sh,0}=v_0/3$ is the ideal speed of an unmodified strong shock in the downstream rest frame. This ensures that upstream turbulence has enough space to develop before particles first interact with the shock; test runs injecting farther upstream (e.g., $5000d_i$) yield indistinguishable results.

Motivated by PIC studies indicating electron preheating in the shock foot \cite[e.g.,][]{gupta+24a}, we consider two upstream particle distributions, both defined in the upstream rest frame: (i) a Maxwellian with thermal speed $v_{{\rm th},e}$, and (ii) a monoenergetic distribution. In both cases, the characteristic speed is treated as a free parameter. All production runs use a monoenergetic distribution, with injection speeds specified separately for the 1D and 2D simulations below. This choice concentrates particles near the energies relevant for reflection and acceleration, avoiding a large population of low-energy particles that are advected downstream and primarily add computational cost. To assess sensitivity of our results to initialized distribution, we tested a Maxwellian with thermal speed $v_{{\rm th},e}=50v_{\rm A,0}$, five times the background electron thermal speed ($v_{\rm th,e,0}=\sqrt{m_i/m_e}\,v_{\rm th,i,0}=10v_{\rm A,0}$, assuming equal background electron and ion temperatures). We also varied the monoenergetic injection speed over a broad range starting at $75v_{\rm A,0}$. Our qualitative conclusions remain unchanged across these tests, provided that the number of macroparticles per cell is large enough to sample both the suprathermal population supplying particles to acceleration and the resulting nonthermal tail.

Over a time interval $\dd t$, the shock sweeps up a gas mass $\dd m=\rho_0\, \dd L_x\, \dd\, L_y \dd\, L_z$ with $\dd L_x=(4/3) v_{0}\, \dd t$. Our main control parameter is the initialized CRe mass supply, $\eta$. For each value of $\eta$, we initialize an upstream test-particle mass $\eta \, \dd m$. We consider five values: $0.05\%$, $0.1\%$, $0.25\%$, $0.5\%$, and $1\%$. These correspond to initialized fractions $0.5 |(q/mc)_e| \eta$ of the background electron charge. Thus, $\eta=0.5\%$ represents a limiting case in which the full background electron charge is initialized in test particles for our chosen electron-ion mass ratio. 

Only a small subset of initialized test particles is ultimately injected into the acceleration cycle and becomes dynamically important. We ``promote'' particles to full CRs (i.e., their current contributes to the feedback terms in the MHD-PIC update) only after they have reached a position $100d_i$ ahead of the ideal shock location\footnote{More advanced ``promotion'' prescriptions based on the particle reflection and/or interaction with the shock front led to the same result at a moderately reduced computational cost.}.

We first present one-dimensional runs with resolution $\Delta x = 1d_i$ and longitudinal domain size $L_x=38{,}400d_i$. For particles with $v\simeq 6v_0$, the Larmor radius $\rho_L$ in the unperturbed upstream field $B_0$ is equal to $3d_i$. For two-dimensional tests, we use a domain of size $L_x\times L_y = (38{,}400 \times 400)d_i$ with resolution $\Delta x=\Delta y=1d_i$. We have verified that our conclusions are converged with resolution. 

We express current densities in units of the physical upstream ion current density, $J_0\equiv e n_{i,0}v_{\rm A,0}$, where $n_{i,0}=\rho_0/m_i$. The shock sweeps up mass at the rate $\rho_0(v_0+v_{\rm sh,0})=(4/3)\rho_0v_0$ per unit area. Since the initialized particles carry a fraction $\eta$ of this mass, taking the reflected fraction $f_{\rm refl}$ of the supplied charge flux gives an upper estimate for the current density carried by the initially reflected population:
\begin{equation}
    |J_{\rm CR}| \sim \frac{4}{3} \left|\frac{q}{mc}\right|_e \eta M_A f_{\rm refl} J_0.
\end{equation}
The corresponding fastest-growing Bell wavelength is 
\begin{equation}
\lambda_{\rm Bell}\sim \frac{4 \pi B_0}{|J_{\rm CR}|}\sim
\frac{3\pi}{|q/mc|_e\eta M_A f_{\rm refl}}d_i.
\end{equation}
For $|q/mc|_e=100$, $M_A = 50$, and a typical reflected fraction $f_{\rm refl} \sim 0.2$, this gives $\lambda_{\rm Bell} \sim 1-20d_i$ over the range of $\eta$ considered here. Thus, the fastest-growing mode associated with the initial reflected current is only marginally resolved at the largest $\eta$, but is well resolved at the lowest $\eta$. 

In the late-time saturated state, however, the nonlinear Bell evolution rapidly transfers power to larger scales. The saturated state is, therefore, dominated by magnetic structures that are substantially larger than the initial fastest-growing linear mode. These larger structures are also the most effective at scattering the current carrying particles, whose Larmor radii are comparable to the fluctuation scale. Consistent with this picture, our higher-resolution tests reach the same saturated magnetic energy and particle spectra, although the time required to reach saturation changes modestly.

For diagnostics, we determine the actual shock position, $x_{\rm sh}(t)$, by locating the outermost position where the smoothed gas velocity crosses $-v_0/2$ and remains decelerated over the next $250\,d_i$ downstream. The resulting $x_{\rm sh}(t)$ is averaged across five outputs and used for the coordinate $x-x_{\rm sh}(t)$ in the diagnostics. Since the shock advances through intermittent reformation steps, we measure the shock speed $v_{\rm sh}$ as the slope of a linear fit to $x_{\rm sh}(t)$ over $100\,\Omega_{\rm ci}^{-1}$. Particle initialization and feedback activation use the prescribed position $x_{\rm sh,exp}(t)$ defined above.

\section{Results}\label{sec:results}
{In this section, we present results of MHD-PIC simulations of non-relativistic, quasi-parallel shocks with an upstream population of preheated electrons. The main result is the emergence of a common regulated state across a wide range of supplied CRe mass fractions. Once nonlinear current regulation is established, increasing the particle supply primarily changes the approach to saturation and the low-energy particle distribution, while the saturated magnetic amplification, far-upstream current, and high-energy spectral normalization depend only weakly on it. We first examine how this state develops in 1D, and then extend our results to 2D runs. We conclude by describing the convergence studies. }

\subsection{Self-regulation via saturation of a Bell-type instability in MHD-PIC}
\begin{figure*}[ht]
    \centering
    \includegraphics[width=\textwidth]{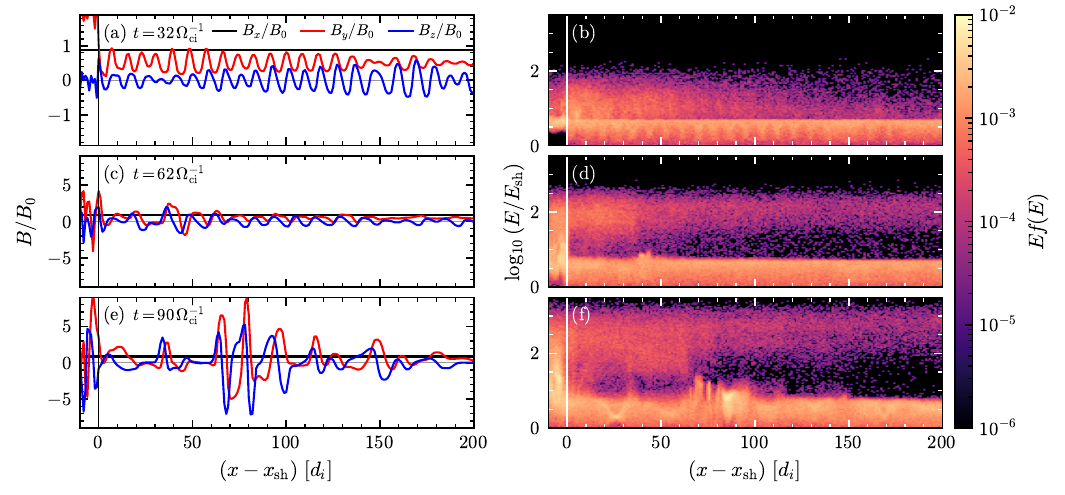}
    \caption{Example of upstream magnetic field evolution and CRe energy phase space for the $\eta=0.25\%$ 1D run. Left column: magnetic field components $B_x/B_0$, $B_y/B_0$, and $B_z/B_0$ as functions of distance from the shock, $x-x_{\rm sh}$, at $t=32\Omega_{ci}^{-1}$ (a), $62\Omega_{ci}^{-1}$ (c), and $90\Omega_{ci}^{-1}$ (e). Right column: corresponding CRe energy phase-space density $E f(E)$, with energy normalized to $E_{\rm sh}$ (b, d, f). The vertical line marks the shock position, shown in black in the left panels and white in the right panels. The upstream turbulence evolves from coherent, gyroradius-scale fluctuations at early times to intermittent, large-amplitude SLAMS in the nonlinear stage, which strongly scatter particles and produce sharp breaks in the upstream phase-space distribution.}
    \label{fig:snapshots_eta25}
\end{figure*}
We introduce CRe upstream as a monoenergetic isotropic distribution of test particles with speed $6v_0$ (measured in the upstream rest frame) and use $\mathrm{ppc}=1000$ macroparticles per cell. We then start depositing their current once they reach a position $100d_i$ ahead of the ideal shock location. A fraction of particles subsequently reflects from the shock and streams back upstream, exciting magnetic fluctuations. Figure~\ref{fig:snapshots_eta25} shows magnetic field profiles (left column) and CRe energy phase space (right column) at three representative times for $\eta=0.25\%$ run. 

At early times, the upstream magnetic perturbations form coherent waves with wavelengths comparable to the resonant scale of the injected particles, $k \rho_L \sim1$, or $\lambda\sim 2\pi\rho_L \approx 18 d_i$ (panel a), where $k$ is the wavevector magnitude and $\rho_L$ is the Larmor radius of the streaming particles. This is consistent with the growth driven by a resonant streaming instability since the CR current is still weak and the non-resonant current-driven mode has not yet become dominant. The polarization of the transverse components supports this interpretation: the wave packets show $B_z$ leading $B_y$, matching the sense of electron gyromotion for reflected electrons propagating away from the shock, from left to right. Panel (b) shows the initial monoenergetic beam together with a reflected, higher-energy upstream population. The initial population is monoenergetic in the upstream rest frame but has a finite energy spread in the downstream rest frame used here.

At later times, the phase relation reverses, with $B_y$ leading $B_z$ (panels c and d). The opposite polarization is consistent with non-resonant current-driven modes, as expected for electron-driven Bell instability \cite[see, e.g.,][]{gupta+21}. As the non-thermal population develops, the upstream current increases and the characteristic $\rho_L$ grows. At the same time, nonlinear Bell evolution transfers magnetic power to larger scales. The upstream turbulence therefore becomes increasingly nonlinear (e and f). 

We interpret the non-linear saturation in terms of current regulation in the Bell instability. When the Larmor radii of the current-carrying CRs in the amplified field become comparable to the dominant turbulent scale, efficient scattering can reduce their streaming relative to the background plasma, weakening the driving current \citep[][]{RiquelmeSpitkovsky2009ApJ...694..626R}. In saturation estimate by \citet[][]{Bell2004MNRAS}, magnetic tension balances the current-driving force, $|\bm{J}_{\rm CR} \times \bm{B}|/c$, on scales of order the CR Larmor radius.

In the nonlinear phase the upstream is dominated by intermittent, strongly compressive Short Large-Amplitude Magnetic Structures (SLAMS; \citealt{SLAMS1991GeoRL..18..373S}; see also \citealt{Hemler2025ApJ...988..144H} and  \citealt{Zekovic2025ApJ...988...40Z} for application to CR diffusion in high Mach shocks) with $\delta B/B\sim 10$ (panels e and f). These SLAMS act as efficient scattering centers: in the energy phase-space map (panel f), they coincide with sharp breaks in the accelerated population, indicating rapid particle scattering. Particles scattered by SLAMS are isotropized and advected downstream, effectively removing them from the acceleration cycle and reducing their contribution to the upstream current. This provides a natural self-regulation mechanism: once sufficiently strong turbulence is generated, the current that drives the non-resonant instability is reduced.

\begin{figure}
    \centering
    \includegraphics[width=\columnwidth]{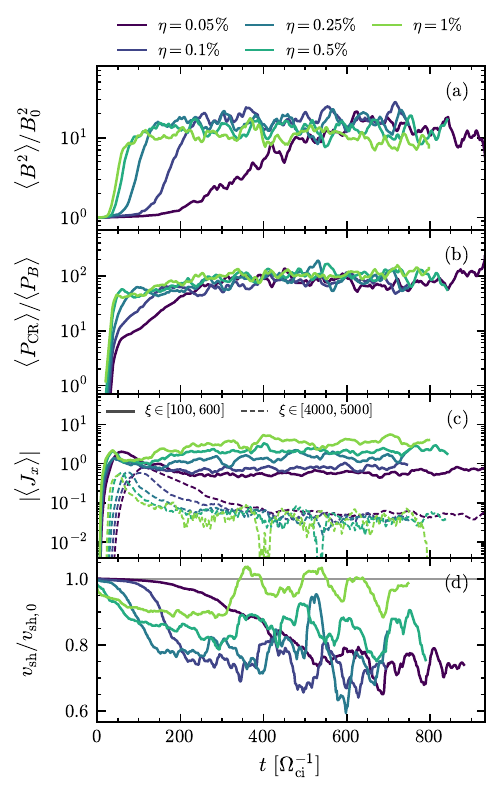}
    \caption{
    Time evolution of (a) the upstream-averaged magnetic energy density $\langle B^2\rangle/B_0^2$, (b) the ratio of upstream-averaged CR and magnetic pressures $\langle P_{\rm CR}\rangle/\langle P_B\rangle$, (c) the magnitude of the upstream-averaged CR current $|\langle J_x\rangle|$, and (d) the shock speed $v_{\rm sh}$ normalized to $v_{\rm sh,0}=v_{0}/3$. 
    Averages are taken in the shock frame, $\xi\equiv (x-x_{\rm sh})/d_i$: solid curves over the near-upstream region $\xi\in[100,600]\,d_i$, and dashed curves in panel (c) over the far-upstream region $\xi\in[4000,5000]\,d_i$. 
    Curves in (a)--(c) are running means over $30\,\Omega_{\rm ci}^{-1}$; $v_{\rm sh}$ is the slope of a linear fit to the shock position over a sliding $100\,\Omega_{\rm ci}^{-1}$ window, omitted in the last $50\,\Omega_{\rm ci}^{-1}$ of each run.
    Colors denote the initial CRe mass fraction $\eta$.
    }
    \label{fig:mag_energy_vel_1d}
\end{figure}
Figure~\ref{fig:mag_energy_vel_1d} shows the time evolution of the upstream normalized magnetic pressure $\langle B^2\rangle/B_0^2$, CR pressure $\langle P_{\rm CR}\rangle / \langle P_B\rangle$, CR current $|\langle J_x\rangle|$, and shock speed  $v_{\rm sh}/v_{\rm sh,0}$ for several initialized CRe mass fractions $\eta$. Despite the order-of-magnitude variation in particle supply, the runs evolve toward a similar saturated state. The upstream magnetic pressure grows rapidly and saturates at $\langle B^2 \rangle /B_0^2\approx 12$ in all cases (a). Increasing $\eta$ primarily reduces the time required to reach saturation. 

For comparison with CR-modified shock models, we compute the scalar, isotropized CR pressure
\begin{equation}
    P_{\rm CR} \equiv \frac{m_s}{3}\int  \gamma_s v_s^2 f({\bm p}_s) d^3p_s.
    \label{eq:pressure}
\end{equation}
Here $f({\bm p}_s)$ is the particle distribution function, $m_s$ is the particle mass, and $v_s$ is the particle velocity measured in the gas comoving frame.\footnote{The CR distribution in the nonlinear state is anisotropic. The isotropized scalar pressure is bounded above by the largest diagonal component. It therefore does not overestimate the maximum stress, although it may underrepresent the component most relevant for momentum exchange along the shock normal.} 
The upstream CR pressure reaches a quasi-steady level earlier than the magnetic pressure and remains larger than the magnetic pressure, $\langle P_{\rm CR} \rangle / \langle P_B\rangle \approx 100$ (b). A large scalar CR pressure does not by itself imply strong shock modification as this depends on the momentum transferred through electromagnetic forces.

Panel (c) shows the upstream CR current, $|\langle J_x\rangle|$. Note that electrons streaming toward $+x$ carry negative $J_x$. In the near-upstream region, $(x-x_{\rm sh})\in(100,600)d_i$ (solid curves), the current increases with $\eta$. However, the scaling is weaker than the linear scaling expected if a fixed fraction of the initialized population remained coherently streaming. This indicates that current regulation already operates close to the shock. Farther upstream, over $(x-x_{\rm sh})\in(4000,5000)d_i$ (dashed lines), the current becomes independent of $\eta$. 

Our interpretation is that the self-generated turbulence scatters and isotropizes the streaming particles as they propagate upstream. Many of these particles are then advected downstream and removed from the current-carrying population. 
Fig.~\ref{fig:mag_energy_vel_1d}c shows that larger $\eta$ initially supplies a larger current, but the far-upstream curves show that the coherent escaping current is regulated to a nearly common value.

The shock speeds remain close to the unmodified value $v_0/3$, consistent with weak hydrodynamic modification, with modest differences among the runs (panel d). The largest initialization fraction $\eta=1\%$ shows a late-time increase in $v_{\rm sh}$ after $\langle P_B\rangle$ saturates. This case corresponds to an unrealistically large CRe supply, exceeding the background electron charge, and should therefore be regarded as an extreme limiting case. Nevertheless, its upstream magnetic pressure, CR pressure ratio, and far-upstream current reach saturation levels comparable to those in the lower $\eta$ cases, indicating that the same current-regulation mechanism still operates. 

\begin{figure*}[ht]
    \centering
    \includegraphics[width=\textwidth]{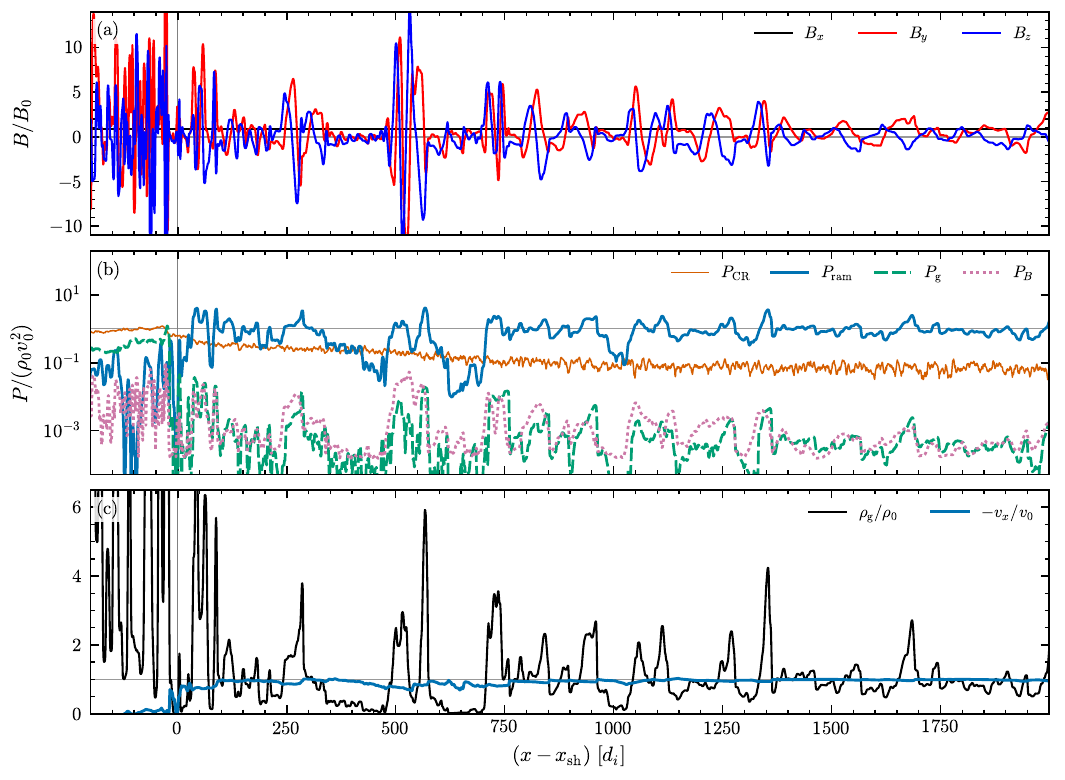}
    \caption{Instantaneous upstream structure of the $M_A=50$, $\eta=0.25\%$ run at $t=234\,\Omega_{\rm ci}^{-1}$, in the saturated state. 
    (a) Magnetic-field components $B_x/B_0$, $B_y/B_0$ and $B_z/B_0$. 
    (b) Ram pressure $P_{\rm ram}=\rho v_x^2$, scalar CR pressure $P_{\rm CR}$, gas pressure $P_{\rm g}$ and magnetic pressure $P_B$, normalized to the far-upstream ram pressure $P_{\rm ram,0}=\rho_0 v_0^2$; $P_{\rm CR}$ is smoothed over $5\,d_i$ to suppress particle noise.
    (c) Gas density $\rho/\rho_0$ and velocity $-v_x/v_0$. 
    In rarefied cavities ($\rho\ll\rho_0$) $P_{\rm CR}$ exceeds $P_{\rm ram}$, while SLAMS, with strong field and high density, dominate the scattering but remain ram-pressure dominated ($P_{\rm ram}\gg P_B$).
    }
    \label{fig:pressure}
\end{figure*}
Figure~\ref{fig:pressure} illustrates the instantaneous upstream structure during the saturated stage for the $\eta=0.25\%$ run. The magnetic field is organized into intermittent, strongly compressive structures (panel a), which coincide with large density and velocity perturbations (panel c). These are the SLAMS that dominate particle scattering in the nonlinear phase \citep[see][]{Zekovic2025ApJ...988...40Z, Hemler2025ApJ...988..144H}.

The pressure decomposition in panel (b) shows that the scalar CR pressure can become locally large. In particular, $P_{\rm CR}$ can locally exceed the ram pressure inside rarefied cavities ($P_{\rm CR}> P_{\rm ram}$ in panel b where $\rho_g\approx0$ in panel c). However, SLAMS that dominate the scattering and momentum exchange remain ram-pressure dominated: $P_{\rm ram}\gg P_{\rm CR}$ in high-$B$ and high-$\rho$ regions. Thus, the instantaneous upstream feedback is highly intermittent and is not well described by a smooth $- \nabla P_{\rm CR}$ force. Instead, the dominant momentum exchange occurs through localized SLAMS.

\begin{figure*}[ht]
    \centering
    \includegraphics[width=\textwidth]{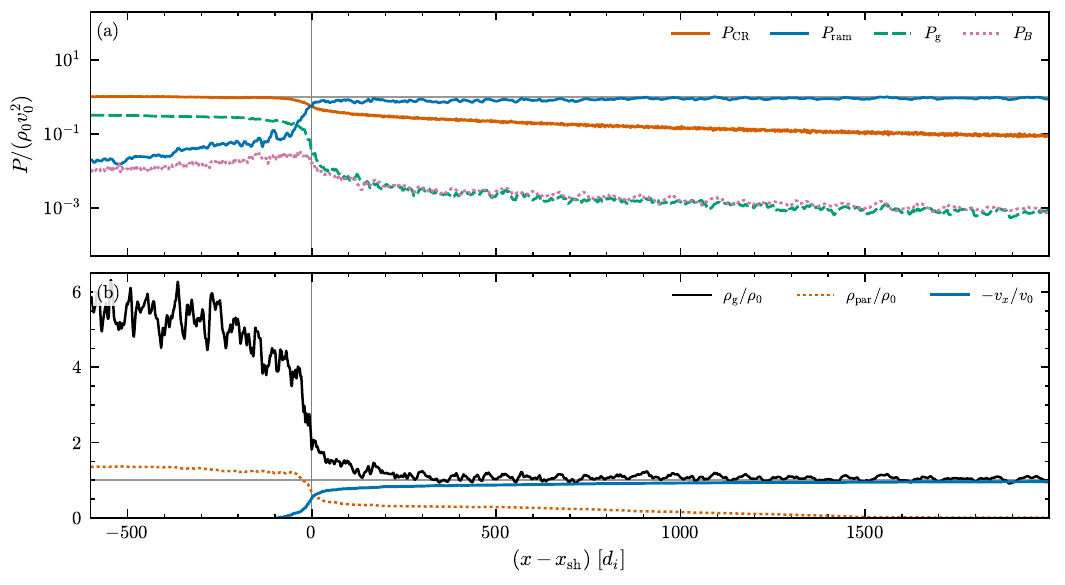}
    \caption{Time-averaged pressure and flow structure of the $M_A=50$, $\eta=0.25\%$ run over $t=200$--$500\,\Omega_{\rm ci}^{-1}$, in the saturated state; profiles are averaged in the shock frame. 
    (a) Ram pressure $P_{\rm ram}=\rho v_x^2$, scalar CR pressure $P_{\rm CR}$, gas pressure $P_{\rm g}$ and magnetic pressure $P_B$, normalized to the far-upstream ram pressure $P_{\rm ram,0}=\rho_0 v_0^2$. 
    (b) Gas density $\rho/\rho_0$, velocity $-v_x/v_0$, and CRe charge density $\rho_{\rm par}/\rho_0$. The vertical line marks the shock position.
    }
    \label{fig:pressure-average}
\end{figure*}

{We next examine how this regulated state modifies the mean shock structure.} Figure~\ref{fig:pressure-average} shows the time-averaged pressure and flow structure in the saturated state. {Despite substantial CR pressure, the flow transition remains localized.} The incoming flow begins to decelerate upstream of the shock, and the density rises smoothly across a localized transition region that extends over a few hundred $d_i$. Although the density is strongly variable because of SLAMS, the downstream density averaged over time and space is above the Rankine-Hugoniot value of $4 \rho_0$, reaching a moderately higher effective total compression of order $5-6$. 

However, this far-downstream gas compression does not necessarily equal to the compression ratio sampled by accelerated particles. The particle density and gas velocity approach their post-shock values over a more limited region, across which the gas density rises to approximately $4.5\rho_0$. The reduced shock speed also indicates a modest increase in the total gas compression: under a steady-flow approximation, $v_{\rm sh}=v_0/(r_{\rm tot}-1)$, so the roughly $25\%$ reduction relative to the unmodified strong-shock speed corresponds to $r_{\rm tot}\simeq5$. Shock deceleration is also found in CR-modified shocks \citep{KangJones2002JKAS...35..159K}.

This distinction is important for interpreting the spectra. In the steady DSA approximation, $q_p=3r/(r-1)$, giving $q_p=4$ for $r=4$ and $q_p=3.75$ for $r=5$. As shown below, we find high-energy spectra close to the test-particle prediction of $q_p=4$, without pronounced concavity over the resolved momentum range. Since the velocity transition is localized, high-energy particles need not sample a strongly momentum-dependent compression ratio. We will describe these spectral trends next.

\begin{figure}
    \centering
    \includegraphics[width=\columnwidth]{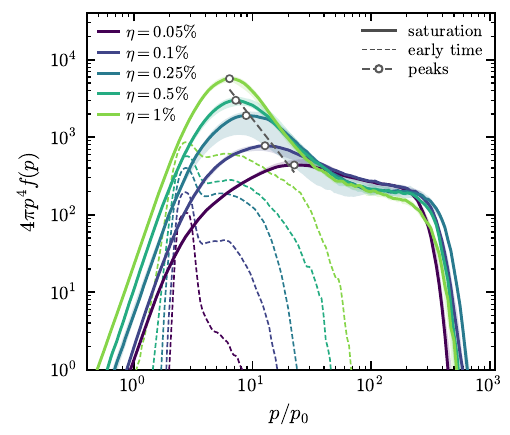}
    \caption{Downstream CRe spectra shown as $4\pi p^4 f(p)$ for 1D runs with different injected mass fractions $\eta$. Each spectrum is computed in a downstream region of $(x-x_{\rm sh})\in(-800,-200)d_i$. Dashed lines show early time spectra at $50\Omega^{-1}_{ci}$, solid lines show median spectra in saturation over a $60\,\Omega_{\rm ci}^{-1}$ window. For each $\eta$ we show a comparable post-saturation dynamical stage, so that the highest-energy cutoff is similar across runs. In the SLAMS-dominated state the spectra develop a pronounced low-energy pile-up and the main peak (indicated by gray circles) shifts systematically toward higher $p$ for smaller $\eta$ (shown by gray dashed line). The highest-momentum tail approaches the test-particle DSA prediction $p^{4}f(p)\approx \mathrm{const}$.
}
    \label{fig:spectra_1d}
\end{figure}

\subsection{Spectral shape}
{The downstream spectra retain a strong dependence on the supplied CRe fraction at low momenta, while their high-energy normalizations converge at late times. 
Figure~\ref{fig:spectra_1d} shows the spectra averaged over $(x-x_{\rm sh})\in(-800,-200)d_i$ at early times and during saturation. At early times (dashed curves, $t=50\Omega_{ci}^{-1}$), the normalization scales approximately linearly with $\eta$, before the non-linear current regulation becomes important.} 

{Once the nonlinear SLAMS regime develops, the low-energy part of the distribution is strongly processed by the self-generated turbulence. Particles with relatively small Larmor radii are efficiently scattered, isotropized, and advected downstream. This produces a distinct low-energy pile-up, or secondary peak. The primary peak, marked by a gray dot for each run, shifts to higher $p$ as $\eta$ decreases (gray dashed line). We attribute this trend to the characteristic scale of the self-generated turbulence. Larger $\eta$ produces shorter-wavelength turbulence, which scatters the low-energy particles more efficiently and returns them downstream after a shorter upstream residence time, giving a comparatively colder processed spectrum. For smaller $\eta$ the turbulence is on larger scales and the scattering is less efficient locally; particles penetrate farther upstream and undergo a more extended sequence of scatterings before downstream advection, producing a comparatively hotter spectrum. This behavior is especially evident in the two-dimensional runs discussed below.}

The low-energy spectral steepening is therefore not a direct NLDSA signature. It may partly reflect the fact that low-energy particles sample an effective compression ratio smaller than $4$, which leads to $q_p>4$. However, this is also the energy range most affected by SLAMS-mediated heating and isotropization, so the low-energy peak and transition region should be interpreted as a processed component rather than as a steady diffusive-acceleration slope.

{CR-modified shock models predict concave CR spectra when particles of increasing momentum sample progressively larger compression ratios. To interpret our spectra in this context,} we compare the width of the time-averaged flow transition with the particle diffusion length, $l_{\rm diff} \sim \kappa / u_1$, where $\kappa$ is the diffusion coefficient and $u_1 = 4 v_0 / 3$ is the upstream velocity in the shock frame. For Bohm-like diffusion, $\kappa\sim \rho_L v / 3$, and Larmor radius $\rho_L = \rho_{L,0} (p/p_0) (B/B_0)^{-1} $, this gives
\begin{equation}
    p/p_0 \sim 2 \sqrt{\frac{l_{\rm diff}}{\rho_{L,0}} \frac{B}{B_0}}.
\end{equation}
From the time-averaged velocity profile (Fig.~\ref{fig:pressure-average}) the size of the precursor is of order of $200d_i$. For $l_{\rm diff}\sim 200d_i$ and $\rho_{L,0}=1 d_i$, this gives $p/p_0 \approx 30-60$ for $B/B_0 \sim 1-5$, which matches the transition range we find in Fig.~\ref{fig:spectra_1d}.

At the highest momenta, SLAMS are less effective at confining particles and the spectrum approaches the test-particle limit ($p^4 f = \mathrm{const}$). This is because a broad precursor extending over the diffusion length of the highest-energy particles is not formed; the hydrodynamic modification is confined to a few hundred $d_i$. The system therefore does not produce the pronounced high-energy concavity expected in NLDSA.

Quantifying an ``acceleration efficiency'' in our closed-box setup is not unique -- a test-particle tail, $f(p)\propto p^{-4}$, implies a formally non-convergent energy integral in the absence of a physical cutoff such as finite age or CR escape. We therefore define a finite-band effective CR pressure,
\begin{equation}
    P_{\rm CR}^{\rm eff} \left( p_{\rm min}, p_{\rm max} \right) \equiv \frac{m_s}{3} \int_{p_{\rm min}}^{p_{\rm max}} \gamma_s v_s^2 f(\mathbf{p}_s) d^3 p_s,
\end{equation}
measured from the downstream non-thermal tail in Fig.~\ref{fig:spectra_1d}. The lower bound $E_{\rm min}$ is chosen above the low-energy thermalized peak, while the upper bound $E_{\rm max}$ is set by the maximum energy reached in the simulations. Assuming approximate isotropy and approximating the high-energy tail by the measured plateau in $p^4 f(p)$, we find $P_{\rm CR}^{\rm eff} \approx 0.3 \rho_0 v_0^2$ over $p/p_0 \simeq 100-1000$.

This quantity should not be interpreted as a global SNR acceleration efficiency. A direct comparison to the $\sim 10-20\%$ energy-budget requirement for Galactic CR production \citep[e.g.,][]{Blasi2013A&ARv..21...70B} depends on the escape time, remnant age, geometry, and (for electrons) radiative losses, none of which are included in our local-box model. The absolute value of $P_{\rm CR}^{\rm eff}$ should be regarded as a diagnostic of the resolved non-thermal tail. {The key result is that the pressure in the resolved nonthermal tail converges to similar values despite the large variation in the supplied CRe mass fraction.}

\subsection{Self-regulation in 2D}
\begin{figure}
    \centering
    \includegraphics[width=\columnwidth]{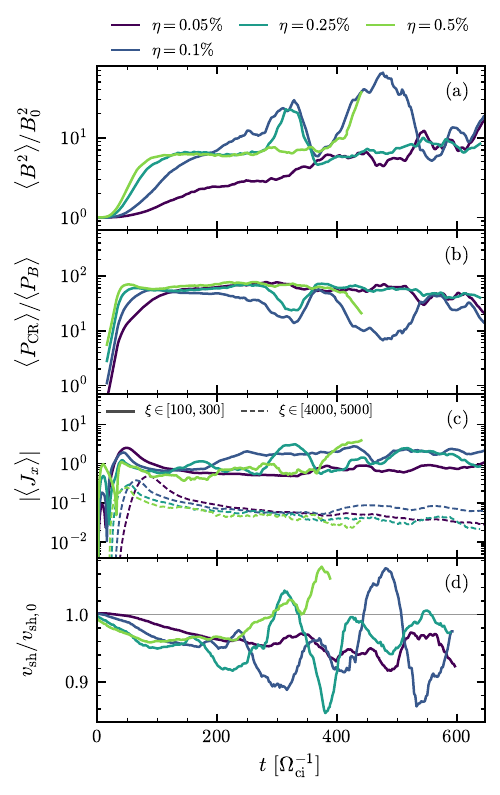}
    \caption{Same as Fig.~\ref{fig:mag_energy_vel_1d} but for the 2D simulations.}
    \label{fig:mag_energy_vel_2d}
\end{figure}
We now demonstrate that the same current-regulation mechanism operates in 2D, where particles and fields have an additional spacial degree of freedom, and the averaging statistics are improved. The transverse box size is $L_y=400d_i$, with periodic boundary conditions in $y$. We consider initialized CRe mass fractions from $0.05\%$ to $0.5\%$. Compared with the 1D runs, we modify the promotion criterion: particles are promoted within $300d_i$ upstream of the ideal shock location, rather than $100d_i$. This is motivated by the broader shock transition observed in 2D. The initialized particles have an isotropic, monoenergetic upstream distribution with velocities $2v_0$ in the upstream rest frame;  we use $\mathrm{ppc}=8$ macroparticles per cell.

Figure~\ref{fig:mag_energy_vel_2d} shows the time evolution of the upstream magnetic pressure, CR-to-magnetic pressure ratio, CR current, and shock speed. Overall, the 2D runs show the same qualitative behavior as in 1D. The upstream magnetic pressure grows and saturates for all $\eta$, reaching $\langle B^2 \rangle / B_0^2 \sim 8$ (panel a), somewhat lower than the 1D saturation level. The pressure ratio $\langle P_{\rm CR} \rangle / \langle P_B\rangle$ (panel b) also approaches comparable saturated values across $\eta$, with values that are modestly lower than in 1D ($\approx 70$ instead of $\approx 100$ in 1D).

Panel (c) shows the upstream CR current, $|\langle J_x\rangle|$. In the near-upstream region (solid lines), the current increases with $\eta$ but the scaling remains sub-linear. Farther upstream, the current is nearly independent of $\eta$ over most of the evolution (dashed lines). As in 1D, we interpret this as evidence that self-generated turbulence scatters and isotropizes the current-carrying particles, limiting the coherent Bell-driving current. We attribute small residual differences between the far-upstream curves, such as the slightly larger current in the $\eta=0.1\%$ run, to intermittency and finite simulation time.

\begin{figure*}[ht!]
    \centering
    \includegraphics[width=\textwidth]{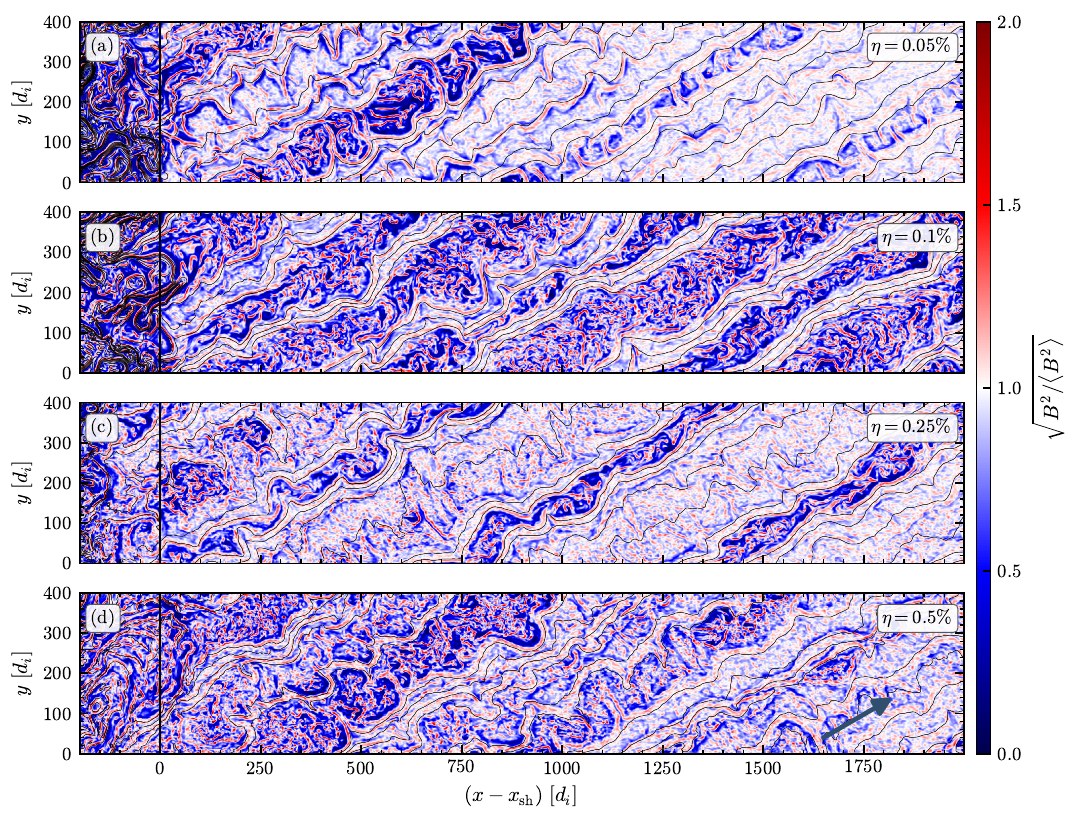}
    \caption{Magnetic field fluctuations, $\sqrt{B^2/ \langle B^2 \rangle}$ (color), and magnetic field lines (black lines) near the shock at a late stage of the evolution, $400 \Omega^{-1}_{ci}$. Panels (a)--(d) show runs with increasing initial CRe mass fraction $\eta$. The arrow in panel (d) indicates the initial magnetic-field direction, inclined by $30 \deg$ to the shock normal.}
    \label{fig:Bfluctuations}
\end{figure*}
Generally, the shock speed $v_{\rm sh}$ decreases modestly in 2D (panel d), consistent with moderate hydrodynamic modification. Short episodes of enhanced shock speed correlate with temporary changes in $\langle P_B\rangle$, $\langle P_{\rm CR} \rangle / \langle P_B\rangle$, and the upstream current, suggesting transient reorganizations of the upstream turbulence and current system. The largest initialization case, $\eta=0.5\%$, shows a late-time shock speed-up. We attribute this to shock reformation mediated by SLAMS. At these late times the shock approaches and eventually overruns the particle-promotion region, so the subsequent evolution should not be interpreted as a physical saturated state. Moreover, $\eta=0.5\%$ is an extreme limiting case, which corresponds to a CRe charge supply equal to the available background electron charge. We, therefore, use this run mainly as a stress test of the current-regulation mechanism. The lower-$\eta$ runs show that magnetic amplification and far-upstream current saturation remain weakly dependent on the initialized particle supply.

Although the saturated upstream magnetic energy is similar across $\eta$, the morphology of the scattering turbulence is different. Figure~\ref{fig:Bfluctuations} shows $\sqrt{B^2/\langle B^2\rangle}$, where $\langle B^2\rangle$ denotes $B^2$ smoothed over a Gaussian kernel of width $5$ cells, and magnetic field lines near the shock at a late time of $400 \Omega^{-1}_{ci}$. At low-$\eta$, (a), the Bell-driven fluctuations organize into coherent channel-like structures, with alternating regions of strong and weak magnetic field. These magnetic channels are anti-correlated with the density:  high-$B$ regions are low-density cavities, while denser channels have weaker magnetic fluctuations. Particles can propagate more easily along the low-$B$ high-density channels, while the high-$B$ structures efficiently scatter particles. 

As $\eta$ increases, the upstream turbulence becomes more volume filling, shifts to smaller scales, and the channel structure becomes less pronounced (d). This morphology enhances scattering near the shock and reduces the fraction of particles that can stream coherently far upstream. Thus, as in 1D, a larger particle supply does not produce a proportionally larger far-upstream current. Instead, the turbulence adjusts to regulate the effective current.

\begin{figure}
    \centering
    \includegraphics[width=\columnwidth]{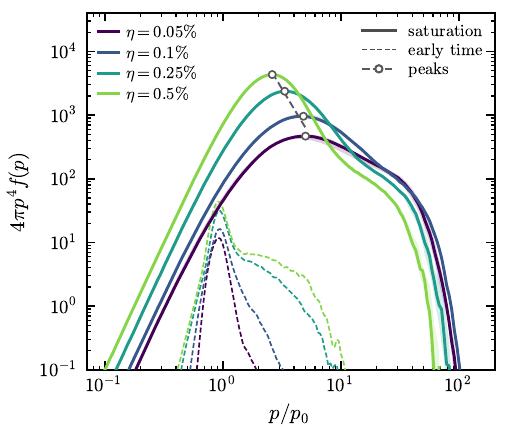}
    \caption{Analogous to Figure~\ref{fig:spectra_1d}, but for 2D runs, with spectra computed in the downstream region $(x-x_{\rm sh})\in[-1500,-700]\,d_i$.
    }
    \label{fig:spectra_2d}
\end{figure}  
Figure~\ref{fig:spectra_2d} shows the downstream CRe spectra in the 2D runs during the saturated stage, measured at comparable late stages for each $\eta$. We average over $(x-x_{\rm sh})\in(-1500,-700)d_i$, sufficiently downstream of the shock to avoid the intermittently broadened shock transition and uncertainties in the instantaneous shock position. 
Overall, the spectra exhibit the same qualitative trends seen in 1D. At early times ($t=50\Omega_{ci}^{-1}$), the normalization scales with $\eta$, and a shock-processed component emerges alongside the initial population. 
At late times, the nonlinear turbulence processes the low-energy particles into a broad peak, corresponding to a heated, partially isotropized component rather than to the initial monoenergetic distribution. As in 1D, the peak positions, marked by open circles, lie at $p/p_0\simeq2.5$--$5$ and shift to lower momenta as $\eta$ increases, consistent with faster and more efficient processing at higher $\eta$. At higher energies, all runs develop a nonthermal tail whose normalization is not very sensitive to $\eta$.

There are also important differences relative to 1D. The maximum energy grows more slowly in 2D; therefore, the resolved high-energy tail spans a smaller dynamic range. Within the present simulation time, the tail remains somewhat steeper than the test-particle plateau, $4\pi p^4 f(p)\simeq{\rm const}$. This does not imply a different asymptotic slope: in the 1D runs, the same momentum range is still part of the transition from the processed low-energy component to the high-energy tail. Longer simulations with larger domains are therefore required to determine whether the 2D tail eventually approaches the test-particle form. Nevertheless, the spectra show that current regulation persists in 2D and that the high-energy normalization converges across $\eta$.

\subsection{Comment on convergence}\label{sec:convergence}
{Beyond varying $\eta$, we test whether the regulated state depends on the prescribed distribution of incoming particles or on the criterion for activating their feedback.} We perform a suite of convergence tests, checking the main diagnostics used in this work: the saturated upstream magnetic pressure, the downstream CRe spectra, and the convergence between different $\eta$ runs. We find that the qualitative conclusions are unchanged across these tests.

We first varied the particle promotion prescription. We tested reflection-based prescriptions, including the maximum upstream displacement relative to the shock and trajectory-based reflection histories, as well as criteria based on kinetic-energy gain during reflection. This affects the early transient evolution and the onset time of saturation, but not the saturated $\langle P_B\rangle$, the high-energy downstream spectra, or the convergence for different $\eta$.

We also tested different initial particle distributions, including thermal distributions and monoenergetic distributions with different injection energies.  These choices primarily change how many particles can reflect and enter the acceleration process. For example, increasing the monoenergetic injection energy or using a distribution with a larger suprathermal tail increases the effective reflected supply. Therefore, changing the initial distribution mainly renormalizes the effective injection fraction. Once the nonlinear Bell-regulated state is reached, the saturated upstream magnetic pressure and the high-energy part of the downstream spectrum converge to the same values. 

Varying the deletion distance for particles leaving the computational domain does not change the saturated state, although placing the boundary close to the shock produces a slight additional shock deceleration, consistent with a small cooling effect from removing energy. Changing when and where test particles are introduced, including varying the initialization distance from $1000d_i$ to $5000d_i$ and delaying initialization until $300 \Omega_{ci}^{-1}$, shifts only the onset of saturation. 

Increasing the number of macroparticles per cell reduces statistical noise in the spectra and spatial profiles, but does not change the mean saturated magnetic pressure or the downstream spectral shape, including the processed low-energy peak. Higher spatial resolution gives the same saturated magnetic energy and spectra, although the time required to reach saturation changes modestly because the earliest Bell growth is better resolved. Changing the magnetic-field inclination from $20^\circ$ to $30^\circ$ also leaves the overall evolution unchanged. 

Finally, we varied MHD-PIC parameters. Reducing the artificial speed of light from the fiducial value of $10^4$ to $10^3$ or $250$ lowers the maximum particle three-velocity and therefore limits the CR current. As expected, the Bell instability then saturates at a lower $\langle P_B \rangle$. Nevertheless, the same current-regulation mechanism persists. We also tested different charge-to-mass ratios and found qualitatively similar behavior. A detailed study of CR ions will be presented in a companion paper.

\section{Discussion}\label{sec:discussion}
Our simulations of high-Mach-number quasi-parallel shocks do not develop a broad, smoothly decelerating precursor associated with the classical NLDSA picture. This is because the strong CR pressure gradient is concentrated in the region of active SLAMS close to the shock.

We interpret this localization as confinement by SLAMS that develop from CR-driven Bell turbulence. The weak dependence of the far-upstream current on the supplied CRe fraction suggests that additional particles remain trapped near the shock while a similar current leaks farther upstream. This regulated transport limits the spatial extent of the CR-pressure buildup. Momentum transfer occurs mainly through interactions with individual SLAMS near the shock, but its time-averaged effect remains consistent with the mean CR-pressure gradient.

The entropy proxy $P_g/\rho^\gamma$ also rises toward the shock, indicating nonadiabatic heating, which we associate with dissipation in the turbulence and SLAMS. Together with CR-induced deceleration, this heating reduces the sonic Mach number entering the subshock \citep[see also][]{Caprioli2009MNRAS.395..895C}, although the subshock remains strong.

In this picture, upstream confinement by Bell-generated SLAMS concentrates the pressure buildup, momentum deposition, and heating near the shock. Despite substantial CR pressure, magnetic amplification, and a total compression ratio $r_{\rm tot}\simeq5$, the transition remains localized and the nonthermal spectra show no pronounced concavity over the resolved range. This mechanism differs from postcursor steepening, where downstream scattering-center drift reduces the effective compression sampled by accelerated particles \citep{HaggertyCaprioli2020ApJ...905....1H, DiesingCaprioli2021ApJ...922....1D}. We plan to develop a one-dimensional NLDSA model in which confinement and leakage through SLAMS are represented by an effective diffusion coefficient that depends on position and particle energy. Coupled with upstream heating and informed by these MHD-PIC simulations, this model will test whether SLAMS-regulated transport can account for the localized precursor and absence of pronounced spectral concavity.

\section{Summary and Conclusion}\label{sec:summary}
We have used 1D and 2D MHD–PIC simulations of non-relativistic, quasi-parallel, high-Mach-number shocks to follow the long-term nonlinear coupling between CR electrons, self-generated turbulence, and the MHD shock structure. The thermal plasma is evolved as an MHD fluid, while CR electrons are advanced kinetically and exchange momentum and energy with the gas. We focus on CR electrons because their larger charge-to-mass ratio allows a stronger driving current than protons at the same CR rest-mass fraction and streaming velocity. While the large electron mass fractions supplied here are not intended to represent realistic astrophysical injection efficiencies, they allow us to explore an extreme regime of current-driven shock regulation. The present study should therefore be interpreted primarily as a study of current-driven electromagnetic feedback and its saturation, rather than as a complete model of ion-dominated SNR CR acceleration. Future work will demonstrate that a similar regulation mechanism operates for CR ions.

{Our main result is that the saturated state depends only weakly on the initialized CRe mass fraction $\eta$. Across the range of $\eta$ explored here, the upstream magnetic pressure and far-upstream CR current converge to similar saturated levels (Fig.~\ref{fig:mag_energy_vel_1d}). Increasing $\eta$ mainly reduces the time required to reach saturation and changes the low-energy particle processing. }

{We interpret this convergence in terms of Bell-driven current regulation. Reflected CRe carry an upstream current that drives non-resonant Bell turbulence. In the nonlinear state, the turbulence forms large-amplitude compressive structures that scatter and isotropize the current-carrying particles. This reduces the coherent upstream current and limits further magnetic amplification,  preventing a larger supplied CRe population from producing a proportionally larger coherent far-upstream current.}

The effective CR pressure in the resolved nonthermal tail is substantial, $P_{\rm CR}^{\rm eff}\sim0.3\rho_0v_0^2$ over $p/p_0\simeq100 - 1000$ and is independent of $\eta$. We treat this quantity as indicative of the maximum acceleration efficiency in the resolved energy range, since the simulations do not include full geometry, escape, remnant evolution, or radiative losses.

Although the acceleration efficiency is high, the shock is not strongly modified as in the classical NLDSA model. The time-averaged flow develops a smooth localized transition in hydrodynamic quantities across the shock (Fig.~\ref{fig:pressure-average}). The overall compression only moderately exceeds the Rankine--Hugoniot value of $4$. The compression sampled by accelerated particles stays close to $4$ and is momentum-independent at high energies. As a result, the 1D high-energy spectra approach the test-particle limit of $4\pi p^4 f(p)\simeq{\rm const}$ over the resolved momentum range. We do not find the pronounced high-energy concavity expected if the particles sampled a broad, smoothly decelerated precursor. The low-energy component is strongly processed by the upstream turbulence.

We find that the same current-regulation mechanism operates in 2D. The upstream magnetic energy and far-upstream current are again not very sensitive to $\eta$ (Fig.~\ref{fig:mag_energy_vel_2d}). The morphology of the 2D turbulence changes with $\eta$: low-$\eta$ runs show coherent channel-like structures, while higher-$\eta$ runs develop more volume-filling turbulence that is more efficient in disrupting coherent streaming of particles (Fig.~\ref{fig:Bfluctuations}). The 2D spectra confirm convergence of the high-energy normalization, but the asymptotic high-energy slope remains unresolved (Fig.~\ref{fig:spectra_2d}).

{We show that current regulation can drive high-Mach-number quasi-parallel shocks toward a common saturated state despite large variations in the supplied CRe population. Substantial CR pressure coexists with a localized flow transition and spectra without pronounced concavity over the resolved momentum range. }These results motivate testing how effectively diffusion closures represent intermittent transport through SLAMS. Future CR-ion, multi-species, and fully 3D simulations will be needed to determine the quantitative saturation level, acceleration efficiency, and asymptotic spectral slope for astrophysical SNR shocks.

\acknowledgments
AG thanks Rui Xu for useful discussions. This research was supported by NSF through Multimessenger Plasma Physics Center (MPPC, NSF grants PHY-2206607 and PHY-2206609), NSF grant AST-2510950, and by Simons Foundation (grant MP-SCMPS-0000147). In addition, this research was supported in part by NSF grant PHY-2309135 to the Kavli Institute for Theoretical Physics (KITP). The computations in this work were, in part, run at facilities
supported by the Scientific Computing Core at the Flatiron Institute, a division of the Simons Foundation. The work was performed using the Princeton Research Computing resources at Princeton University which is a consortium of groups led by the Princeton Institute for Computational Science and Engineering (PICSciE) and Office of Information Technology’s Research Computing.
\bibliography{ref}

\end{document}